# Exploring Mindset's Applicability to Students' Experiences with Challenge in Transformed College Physics Courses

Angela J. Little,[1] Bridget Humphrey,[2] Abigail Green,[3] Abhilash Nair,[1] and Vashti Sawtelle[3,1]

[1]*Department of Physics and Astronomy, Michigan State University, East Lansing, MI 48823, USA*
[2]*Department of Neuroscience, Michigan State University, East Lansing, MI 48823, USA*
[3]*Lyman Briggs College, Michigan State University, East Lansing, MI 48823, USA*

The mindset literature is a longstanding area of psychological research focused on beliefs about intelligence, response to challenge, and goals for learning (Dweck, 2000). However, the mindset literature's applicability to the context of college physics has not been widely studied. In this paper we narrow our focus toward students' descriptions of their responses to challenge in college physics. We ask the research questions, "can we see responses to challenge in college physics that resemble that of the mindset literature?" and "how do students express evidence of challenge and to what extent is such evidence reflective of challenges found in the mindset literature?" To answer these questions, we developed a novel coding scheme for interview dialogue around college physics challenge and students' responses to it. In this paper we present the development process of our coding scheme. We find that it is possible to see student descriptions of challenge that resemble the mindset literature's characterizations. However, college physics challenges are frequently different than those studied in the mindset literature. We show that, in the landscape of college physics challenges, mindset beliefs cannot always be considered to be the dominant factor in how students respond to challenge. Broadly, our coding scheme helps the field move beyond broad Likert-scale survey measures of students' mindset beliefs.

## I. Introduction

The mindset literature centrally focuses on beliefs about intelligence. If a person believes that it is possible to grow and improve their intelligence or talent, research suggests that this helps them in their educational endeavors and through moments of challenge and failure. Underlying mindset is a socio-cognitive theoretical framework. According to this framework, people's experiences with challenge are filtered through their beliefs, making it likely that a person will respond in more or less productive ways (Dweck, 2000). Two major concerns about the mindset literature's applicability to college physics include the following: one, only limited types of challenge have been studied in the mindset literature and, two, beliefs about intelligence have primarily been examined through survey methodologies. In this paper, we present the development of a novel coding scheme that engages with these concerns, ultimately raising important questions about the mindset literature's applicability to transformed college physics courses.

The theoretical framework underlying mindset studies was built on research that only examines a limited set of challenges (e.g. Diener & Dweck, 1978; Diener & Dweck, 1980). These challenges are typically characterized by close-ended answers (e.g. logic problems), near certain possibility of failure, short time periods of a few hours, and an individual performing the activity alone. Challenges studied in the mindset literature are quite different from more active-learning and project-based college physics classrooms, *transformed* classrooms, where students are frequently given open-ended problems where success is possible, work on projects over many weeks, and engage with their peers to solve problems. The differences in the characteristics of challenge across mindset studies and transformed college physics courses raises questions: Is the theoretical framework employed in mindset studies applicable to transformed college physics? Are the mechanisms behind different student responses to challenge best explained by students' beliefs about intelligence? The answers to these questions are consequential to educators considering mindset interventions in college physics.

The mindset literature's survey methodologies treat beliefs about intelligence as somewhat stable and binary. Research participants are typically described as having either a *growth* or a *fixed* mindset. A growth mindset refers to believing it is possible to improve one's intelligence and responding productively in the face of challenge. A fixed mindset refers to believing that it is not possible to improve one's intelligence and responding less productively in the face of challenge (Dweck, 2000). Dweck herself recently remarked, "First, we thought people had one mindset or another…we still often assess mindsets that way, but we've discovered over time there are so many triggers in the environment that put any of us into more of a fixed mindset" (Sparks, 2017). As Dweck notes, the mindset literature has not moved beyond survey-use for the measurement of intelligence beliefs. In this paper we will present a novel methodology by which researchers can look for context-dependent variation in mindset beliefs. This methodology examines student reflections about challenge in interview data.

What kind of mindset-related variation in student dialogue might be important for researchers and educators to understand? To exemplify one kind of variation, we provide a short excerpt of an interview we performed with a biology major named "Leyla." Leyla is enrolled in a transformed introductory college physics course. Near the end of the course, she compares herself to her peers and deems herself not good at physics: *"That's why I think I'm bad at physics. Because it does not, in any way, come easy to me…some of my peers…they'll see a problem, they just know how to do it, and I don't…know how they can do that. It's just so hard for me."* This dialogue appears aligned with a fixed mindset: Leyla encounters challenge and makes a negative statement about her ability in physics. Yet, in this same interview, Leyla describes working incredibly hard in response to challenging course and project-based activities. For instance, Leyla notes, *"I came in thinking, just, 'I'm going to go to all the office hours, just going to study, do what I need'…I would come into office hours, and I was like, 'Okay. I understand it… I can apply it to any question.' So, I came in not knowing it, and then…Okay, it's really - you see it happening. It's not just on paper."* Here, Leyla describes working hard and seeking help through office hours, a description of behavior that is aligned with a growth mindset. Leyla's example shows the importance of not assuming that student dialogue will neatly fit entirely into growth or fixed mindset categories. A student might report that they are "bad at physics," but work hard and be willing to show vulnerability to their instructors in the face of challenge.

This paper focuses on what can be learned from the development of a novel coding scheme to capture variation in student interview dialogue around challenge in college physics. To be clear, this paper is focused deeply on the coding development itself. In the process of asking the research question, "can we see mindset-related variation in students' descriptions of their responses to challenges in a college physics course?" many theoretical and methodological insights were gained. Other researchers have argued that important lessons can be learned by sharing and reflecting on the development process of research methodologies (Schoenfeld, 1992; Hammer & Berland, 2014). In this paper we present two distinct coding scheme development processes: coding for challenge and coding for response to challenge.

**II. Overview**

Our interest in understanding how students respond to challenge requires beginning with understanding challenge itself. We first describe how we find evidence for challenge in student dialogue. After showing how our codes evolved over time, we present the finalized codes. We then discuss the implications of our codes. In this discussion we make two main points: (1) transformed college physics challenges are distinct from challenges studied in the mindset literature; and (2) the mindset literature's approach of treating all challenges the same way theoretically will likely prove problematic in transformed college physics.

We then describe our coding scheme development for how students respond to challenge. We describe how the mindset literature categorizes response to challenge and how this led to a set of *a priori* codes. We then describe how those codes evolved over time when examining college physics interview data. In the discussion of the implications of our codes we will make two main points: (1) our codes can be used by researchers to take a context-dependent belief approach to studying mindset; and (2), our codes must be used with consideration of students' broader educational context. We explore how a number of competing explanations exist, beyond intelligence beliefs, for why a student might respond to challenge in ways aligned with a particular mindset. Ultimately, our methodological development raises questions not only about treating intelligence beliefs as binary, but of beliefs more generally as the best explanatory model for how students respond to challenge in college physics.

### III. Literature Review
Mindset is a longstanding area of psychological research that examines beliefs about intelligence, goals for learning, and responses to challenge. The mindset literature tends to categorize students as having either fixed or growth mindsets: Students with a fixed mindset (FM) "*believe that their intelligence is something that is finite and unchangeable. This makes them doubt their intelligence when they experience difficulty and it undermines resilience in learning.*" In contrast, students with a growth mindset (GM) "*believe that intelligence can be developed. In this mindset, students respond more resiliently to challenges and show greater learning and achievement in the face of difficulty*" (Yeager et al., 2013). Growth and fixed mindset are umbrella terms that encompass certain types of beliefs, goals for learning, and responses to challenge. We will sometimes use these umbrella terms to refer to any of these three areas unless a more specific construct is needed for clarity.

Mindset research is vast and encompasses nearly fifty years of scholarship. Summaries of the mindset literature are available (Dweck, 2000; Yeager & Dweck, 2012; Yeager et al., 2013). In this paper, we will make an argument that narrowing our focus to understanding how people respond to challenge will provide insight into the applicability of mindset research in college physics. Learning goals, another central aspect of the mindset literature, will not be addressed in this paper as to keep a focus on response to challenge. In this literature review, we first characterize the basic elements of how mindset is studied. Then, we show that college STEM, particularly college physics, is an understudied area of mindset research. We characterize how mindset research has mainly focused on beliefs and not challenge when considering equity issues. We end this section with a description of the theoretical frameworks around epistemology in college physics that informed our methodological development.

A. Methodologies of Mindset Research

To measure beliefs about the nature of intelligence, the Implicit Theory of Intelligence Survey (Dweck, 2000) was designed to measure entity theory (fixed mindset) and incremental theory (growth mindset) beliefs. It involves statements like "your intelligence is something about you that you can't change very much" and asks people to rate their agreement on a Likert scale. Researchers will sometimes create surveys that add "in math" at the beginning of each statement to probe math-specific intelligence beliefs (Good et al., 2012). This is the most context-specific way that psychology researchers have studied beliefs.

Mindset research takes a focus on intelligence beliefs because of seminal studies dating back to the 1970s and 80s. A typical early mindset study would involve bringing children to a campus psychology lab, exposing them to a challenge, and then measuring how they responded to that challenge (Diener & Dweck, 1978). For instance: would a child give up in the face of a challenging logic problem or would they keep trying? At the time, researchers were surprised that children's past performance or grades in school were not actually predictive of how they responded to challenge. Rather, children's beliefs about intelligence were most predictive of their responses to challenge. For instance, some children quickly gave up while other children continued working in the face of failure and difficulty (Dweck, 2000). These studies solidified the applicability of mindset's socio-cognitive theoretical framework for understanding children's responses to challenge. Beliefs were treated as binary and fairly stable, but possible to impact. Relationships between beliefs, response to challenge, and goals for learning were explored across many studies.

After showing that children's intelligence beliefs were important to the process of learning in the clinical lab setting, researchers worked on understanding mindset in real-world contexts. This has included work linking mindset to course performance and interventions to change mindset (Dweck, 2006; Dweck, 2008; Yeager et al., 2013). Mindset interventions have been designed with the goal of positively impacting student success. Interventions in which students learn about the plasticity of the brain have had remarkable success in improving math grades and college persistence, particularly for populations of students who deal with racial stereotypes about their intelligence and/or who are struggling with grades (Aronson et al., 2002; Blackwell et al., 2007; Yeager et al., 2013; Yeager et al., 2016). However, very little is understood about the mechanisms by which these interventions seem to work (Schwartz et al., 2016). One of the most recent large mindset intervention studies (Yeager et al., 2016) raised some important questions about mechanism. Somewhat perplexingly, the intervention supported students with lower grades to improve them, however, there was no measurable change in mindset beliefs with these students who had improved their grades. In addition, some of the better performing students did experience mindset belief change but their grades did not improve. Schwartz et al. (2016) suggest that social-psychological interventions in mindset, as well as a few other areas, present "a grand challenge for the field…to develop better instrumentation that can capture relevant behaviors and attitudes over time, and how these vary across context and population." Recent work has attempted large-scale interventions, but there are very few qualitative studies that shed light on mechanism.

Some newer research, although not qualitative in nature, has worked to understand aspects of learning environments with respect to mindset. For instance, researchers have characterized teachers' self-reported instructional practices and their connections with mindset beliefs. In the context of early grade school, it was found that fixed mindset-related instructional practices affected student mindset (Park et at., 2016). Likewise, researchers have shown that the type of praise that parents give to children can affect the children's mindsets (Pomerantz & Kempner, 2013).

B. Mindset and College STEM

Although mindset is a longstanding area of psychological research, college STEM, particularly college physics, is not a central focus. Only a handful of studies on mindset have been performed in STEM at the university level. The studies that exist focus almost entirely on introductory calculus courses (Good et al., 2012; Rattan et al., 2012) with some studies on introductory computer science courses (Lewis et al., 2011, Flanigan et al., 2015) and general chemistry (Grant & Dweck, 2003). Two studies exist in physics and they focus on post-college years: one is a study about Ph.D. publication records in physics and chemistry (Hazari et al., 2010) and the other narrows into physics faculty thinking about graduate admissions (Scherr et al., 2017). At the time of submission of this paper, there were only sixteen articles in the PER-focused journal within *Physical Review* that cite a study by Dweck. Only one of these sixteen articles (Scherr et al., 2017) explicitly focuses their research program on mindset, but, again, the focus on this research was on faculty and not students. The most detailed study around mindset with students in college physics is a Masters thesis showing that having a growth mindset correlated with higher Force Concept Inventory gains in a Modeling Instruction context at a large research university (Megowan-Romanowicz, 2011). In terms of Schwartz et al.'s 2016 call for more detailed methodologies to understand mechanisms behind mindset, college STEM does not have any such methodologies. Rather, mindset is broadly researched as a factor in student success.

Although not formally framed as work on mindset, some standard Physics Education Research survey tools do probe nearby belief structures. These tools do not explicitly reference the mindset literature. For instance, the Epistemological Beliefs Assessment for Physical Science (EBAPS) has a cluster referred to as "source of ability to learn." This cluster includes Likert scale statements such as, "Someone who doesn't have high natural ability can still learn the material well even in a hard chemistry or physics class" (Elby et al., n.d.). Similarly, the Colorado Learning Attitudes about Science Survey (CLASS) has a few questions under a Sense-making/Effort cluster such as, "Nearly everyone is capable of understanding physics if they work at it" (Adams et al., 2006). The experimental physics version of the CLASS survey (referred to as the E-CLASS) similarly has questions such as, "If I try hard enough I can succeed at doing physics experiments" (Zwickl et al., 2014). Likewise, the Views about Science Survey (VASS) has Learnability cluster that probes whether science is learnable "by anyone willing to make the effort" (Halloun & Hestenes, 1998). Taken together, the sets of questions on these surveys could be considered initial work in understanding mindset in more physics course-specific contexts. However, clusters or questions related to mindset are not often explicitly studied in research on these surveys. Additionally, researchers have not explicitly mapped out the connections (or lack thereof) between these survey items and mindset's theoretical framework and methodologies.

C. Mindset Research and Equity

A number of research studies have shown that mindset interventions appear to have a stronger effect on students who are part of racial or gender stereotyped groups with respect to intelligence and who are also struggling with respect to grades (Aronson et al., 2002; Blackwell, 2007;

Yeager, 2016). In her book, Dweck (2000) discusses the possibility that mindset interventions might reduce stereotype threat, a condition in which stereotyped groups perform worse under high stress testing conditions. She describes a study by Aronson & Fried:

*"Aronson reasoned that the climate of a predominately white college adds an extra burden for African American students...Aronson and Fried (1998) decided to see whether teaching undergraduate students an incremental theory would reduce stereotype threat and improve their college performance…"*

The Aronson and Fried (1998) study showed that a mindset intervention study "appreciably reduced" a Black-white "achievement gap"[1] (measured by grades) seen in the control group at a selective predominately white university. Likewise, Good et al. (2012) showed that a growth mindset acted as a kind of buffer against factors such as gender stereotypes that eroded female students' sense of belonging in college math.

Research at the intersection of equity and mindset has typically considered intervening with student beliefs rather than with the environment itself. Aronson and Fried's (1998) work, as well as work on stereotype threat more generally (Steele & Aronson, 1995; Spencer et al., 1999) highlights an important consideration: racism and sexism can impact *how students experience challenge* such as a high stakes mathematics exam. Some students might experience stereotype threat, for instance, while others do not. This directs us to pay attention to how student experiences with marginalization may impact their experiences of challenge. In addition to stereotype threat effects in test-taking situations, racial and/or gender stereotyped students experience many other effects of racism and sexism in day-to-day classroom environments in college STEM (Fries-Britt & Turner, 2001; Fries-Britt & Turner, 2002; Fries-Britt et al., 2010; Harper, 2013; Harwood et al., 2015; Northwestern University, Black Student Experience Task Force, 2016; Rosa & Mensa, 2016; McGee & Bentley, 2017). More work is needed in this arena with respect to mindset studies.

D. Context-Dependent Beliefs in College Physics Research
Context-dependent approaches to measuring student mindset in college STEM are currently limited. Although Dweck herself (2015) has called for a less binary approach in how mindset beliefs are studied, research methodologies have not yet caught up. Researchers in college computer science courses have used discipline-general survey measures of mindset (Flanigan et al., 2015). Researchers in college math have developed surveys that ask about math-specific mindset beliefs (Good et al., 2012). We argue for building even more detailed context-specific measures that help us understand the mechanisms behind the role of mindset in college physics.

To build context-dependent mindset measures, we draw upon lessons learned from the physics and science beliefs research literature. Researchers in college physics, for instance, have developed a number belief surveys (e.g. MPEX, CLASS, etc.) that treat beliefs about physics as fairly stable (Redish et al., 2000; Adams et al., 2006). However, the field has also benefitted from "naturalistic case studies, including open-format interviews" that explore context-dependency (Hammer & Elby, 2002). Such case studies and interview methodologies have

---
[1] For critiques of achievement-gap framing see (Ladson-Billings, 2006)

demonstrated ways in which students can display different kinds of beliefs. For instance, Elby and Hammer (2010) give the following example of a college physics student:

*"When asked how he prepared for his physics test, Louis said that he "studied every word of those homework solutions...I was memorizing the book, too." This response reflects a view of knowledge as something absorbed from an authoritative source. By contrast, talking about his strategies for tutoring other students, Louis said, "what I like to do is build on what they already know instead of introducing a totally new concept," reflecting a view of knowledge as something constructed out of prior knowledge. Louis's variability can be understood as arising from the different contexts activating different resources, in this case resources for understanding the nature of knowledge and how an individual comes to have it"*

Researchers have adopted the language of "epistemological resources" to highlight the varied ways that students can reason about the nature of knowledge (Hammer & Elby, 2003; Hammer et al., 2005; Elby & Hammer, 2010). Most notably, context-dependent beliefs are not studied through surveys. Rather, student dialogue about *one's actions in physics classes* are treated as representing certain context-dependent beliefs. We drew on this theoretical framework in the data collection design underlying our coding scheme development.

Two recent studies suggest that coding schemes that allow for seeing variation in mindset-related dialogue hold promise. In prior work (Little at al., 2016), we built a more general mindset-related coding scheme to qualitatively examine student interview dialogue across the topics of family stories, K-12 experiences, and college physics. We showed preliminary evidence of variation in mindset-related talk across intelligence beliefs and learning orientations. Likewise, Scherr et al.'s (2017) complementary work on physics faculty dialogue shows that qualitative coding approaches do show variation: faculty interviews can contain a mixture of both fixed and growth mindset-aligned statements.

**IV. Data Collection Methodology**

One of the major research questions underling our methodology was, "can we see mindset-related variation in students' descriptions of their responses to challenges in a college physics course?" To address this question, we began with an interview study with students at the end of one year of introductory college physics. We interviewed five students, refined our interview questions, and then interviewed three additional students the following year. A set of colleagues also shared with us one additional interview data set with thirteen students. As our goal was to develop methodologies through which we could characterize variation, this set of student interviews gave us enough examples to begin with. Limitations of the sample population will be addressed at the end of this paper. More detail about our interview design and student populations are provided, below.

A. Interview Design

One-on-one interviews occurred with students at the end of two semesters of introductory college physics. Interviews were roughly one-hour in length and semi-structured. Drawing on the context-dependent epistemological resources theoretical framework (Hammer & Elby, 2003;

Elby & Hammer, 2010), we designed interview questions that we hypothesized would elicit possible variation in student dialogue about mindset. We crafted questions that we expected would be likely to elicit growth-mindset related talk such as, "Do you feel proud of anything in physics?" This question arose from Little's work suggesting that moments of feeling proud frequently co-occurred with seeing one's growth (Little, 2015). Likewise, we asked questions like, "Did anything seem hard or impossible at first, in physics, but eventually you were able to understand?" Questions that we expected to elicit more fixed-mindset related talk included, "Some people think that to be good at science or physics it has to come easy to you. On the spectrum of agree to disagree where do you put yourself?" Students would be encouraged to explain their reasoning with examples from college physics. Students' responses to these three questions about feeling proud, hard moments, and what it means to be good at science were the predominant responses analyzed in the coding scheme development presented in this paper.

A. Courses and Student Populations Studied
As noted above, we interviewed five students, refined our questions, and then interviewed three additional students the following year. In our recruitment, we focused on classroom contexts of introductory physics for life sciences courses. This meant that students in these courses were primarily biology students, likely to come in with some trepidation around physics, but also willing to put a lot of work into the class (Sawtelle & Turpen, 2016). We also chose a classroom context where these students were most likely to have a *positive* experience. We chose transformed classrooms with optional project-based components. We hypothesized that this population (with some fear of physics, but willing to work hard and in a transformed environment) was most likely to show variation in mindset-related dialogue and was therefore a good starting place.

Recruitment of students focused on two different introductory physics courses for primarily life sciences majors that emphasized interdisciplinary connections, research-based instructional strategies, and project-based components at a single university. A course transformation effort occurred over the years of our study; thus, one course took place in an active-learning large lecture classroom while the other took place in a studio classroom (see table, below). In the first year, we emailed the students who had signed up for an optional project-based component. This was not a large set of students, and five students were willing to participate in our first set of interviews. In the second year, three students were part of an optional project-based component that had a special design to it: a graduate student (Nair) worked with students over an entire semester to design a biomedical device and explore its ethical implications (Vital Signs: Bridging & Democratizing Physics, n.d.). Again, given our focus of talking to students with some fear of physics, but also most likely to have a positive experience with learning physics, this particularly supported group of three students was ideal.

The overall university context was a large, Midwestern research university. Demographic post-surveys were given. Interviewees in the first set of five interviews identified as white (5/5) and predominately female (4/5). In the second set of three interviews, students identified predominately as female (2/3). These students gave the following race and ethnicity responses, respectively: white, with no ethnicity written in; white with Middle Eastern ethnicity; and Asian with Chinese ethnicity.

We also drew on a third interview study performed by colleagues at a small liberal arts college. This interview study took place within a transformed introductory physics course for life sciences majors that included interdisciplinary connections, research-based instructional strategies, and a short (two-week) project-based component. Although interviews performed in this context were not formally designed to probe for mindset, interviews did elicit student dialogue about challenge. We used interviews from this study to have a larger example space. Demographic surveys collected in this interview data set asked for gender information and 7/13 students identified as female. Racial demographics were not collected.

A summary of the course contexts in which we drew interview data from can be found, below:

Introductory Physics for Life Sciences Courses Informing Coding Scheme Development

| Course | Research-Based Instructional Strategies | Institution Type | Optional Semester-Long Project-Based Component? | Number of Interviews | Timing of Interview | Range of Interviewee Grade in First Semester Physics Course |
|---|---|---|---|---|---|---|
| Course 1 | Active Learning, Clicker Questions, Tutorials | Large research university | Y | 5 | End of second semester of college physics | 4.0 |
| Course 2 | Studio Physics, Computational Problems, Modeling Instruction, NEXUS | Large research university | Y | 3 | End of second semester of college physics | 4.0 |
| Course 3 | Clicker Questions, active learning, tutorials, final paper on biological application of physics | Liberal arts college | N | 13 | Middle of first semester or beginning of second semester of college physics | 3.0-4.0 |

**V. Methodological Development and Discussion Overview**
We will present two distinct coding scheme development processes one after the other: coding for challenge and then coding for response to challenge. First, we will describe a priori codes that we developed through examination of the mindset literature. Then, we will describe the process by which these codes were broken apart, clarified, and added to through examining end-of-semester student interview data. We will then present the finalized codes.

We first began with the broad questions, "can we see responses to challenge in college physics that resemble that of the mindset literature?" and "how do students express evidence of challenge and to what extent is such evidence reflective of challenges found in the mindset literature?" These questions were important to answer before being able to make progress on understanding mindset-related variation in students' descriptions of their responses to challenges in a college physics course. This paper focuses on these first two research questions and we will discuss each of them after presenting our coding refinement process and finalized codes. As Engle et al. (2007) point out, much research in the learning sciences takes a kind of top-down meets bottom-up approach. Theory informs how one begins when collecting and analyzing data. Keeping one's research questions general enough allows for discovery of the more unanticipated aspects of the phenomenon that one is studying. Our methodology is consistent with this approach.

## VI. Examining Evidence for Challenge

A. *A Priori* Codes Grounded in the Literature
A goal in this work is to understand the relationship between kinds of challenge in the mindset literature in comparison to challenges experienced in transformed introductory college physics. Therefore, our starting place for coding "challenge" was to describe the ways that participants in mindset studies experienced challenge. The mindset literature does not explicitly define challenge because most studies employ tasks that are constructed to be obviously challenging (e.g. working on problems that are multiple grade levels too high). Thus, we surveyed the literature, making our own characterizations of activities typically featured in mindset studies. We characterize challenges found in the mindset literature the following four ways:

| *A Priori* Code Name | Description |
| --- | --- |
| Isolation | Participants were asked to perform challenges on their own (Diener & Dweck, 1980) |
| Confusing/Dense | Participants sometimes faced challenges that involved making sense of densely written material. Dweck describes a study where students had to read a passage that "was written in a muddy and tortuous style, a style that looked comprehensible on the surface but was quite confusing" (Dweck, 2000). |
| Task Failure | Participants were frequently given conceptual problem-solving tasks that were "too difficult for children their age," resulting in a high probability that they would fail (Dweck, 2000). |
| New Tasks | Dweck notes that students' "intelligence is certainly on the line in settings where they are being evaluated on a host of new intellectual tasks" (Dweck, 2000). In clinical lab studies it is rare that the challenging aspect of a task was described as having to do with its newness. It was almost always so difficult that failure was likely or constructed to be artificially confusing or dense. However, Dweck posited that middle school was a |

| | time of new intellectual tasks, and that this newness was an aspect of challenge that students faced. |
|---|---|

The few psychological studies about mindset in college STEM are not focused on responses to in-the-moment challenges. However, an educational research study in introductory college computer science gave us one additional area of challenge to pay attention to. This study used a grounded theory approach to focus on how students decide to major in computer science (Lewis et al., 2011). One finding from this study was that students frequently compared themselves to their peers and these judgments appeared to be related to mindset. Some students explained their worse performance as due to background; others explained it due to their lack of inherent ability. Lewis et al. characterized this dialogue as related to mindset. We thus included an interactional kind of challenge not present in the psychological clinical research studies: **Peer Comparison-Worse**. We keep peer comparison pulled out of the table, above, to emphasize that peer interaction is not a typical feature of mindset-related studies.

While we started with five *a priori* evidence of challenge codes, we expected that we would find many more kinds of challenge when examining college physics interview data.

B. Selected Methodological Moments: Shifts to Evidence for Challenge Codes
After developing our set of a priori codes, members of our team (Little, Humphrey, and Green) examined multiple sets of student interviews about their experiences with challenge. We flagged student dialogue that fit these broad *a priori* codes as well as anything that seemed related to challenge or students' response to it. We noticed themes when we examined student dialogue that did not fit into an *a priori* code and inductively broke out these themes into additional codes. As an example, one additional code that arose in our data had to do with tasks that were tedious in nature. One student described coding as part of a course project as "tedious." This caused us to articulate a new challenge code that we labeled as **Tedious**. In addition, a number of more interactional challenges arose that had to do with physics course social dynamics not found in individual lab studies.

Examination of student dialogue also led us to modify details of our a priori codes. For instance, the **Task Failure** *a priori* code quickly raised questions in application to our data. Is task failure specific to a student's goals and experiences or something that we as researchers should impose? What grade should count as failing a task? What other markers of failure are there beyond grades? Overall, our approach has been to prioritize *how students themselves* set and evaluate their own goals. We had moments in our data where a student received a B on an exam, but clearly articulated that their goal was an A. We concluded that any clear lack of meeting one's goals should be considered reasonably represented an experience of challenge. In addition, students sometimes describe getting particular homework problems or project tasks "wrong," typically through an instructor giving them feedback. We considered this to also fall under task failure and to be a marker of struggle.

Another methodological conversation that arose had to do with whether our categories were mutually exclusive. For instance, our **Difficult/Hard** code captures fairly general aspects of student language whereas our **Complex/Confusing/Dense** code tries to specifically tag conceptual difficulties. Consider the following student quotes that involve elements of being both **Difficult/Hard** and **Complex/Confusing/Dense**. One student noted that physics is "*It's just so hard to understand*" and another student described a "*really hard concept that I don't get.*" Ultimately, we decided that our **Difficult/Hard** code could reasonably act as a broad code that may co-occur with other codes. We made this decision in part because we wanted to capture as much information as possible about student experience with challenge.

The above process of modifying a priori codes in conversation with our data led to an intermediate set of codes with names and descriptions. Our next step was to perform an activity practiced in law (Rissland, 1983) where we took student dialogue examples that fit our codes and imagined modifications to those examples. We then asked: would that modified example fit? In other words, we developed hypothetical examples that we felt should fit our codes but that the language of our codes did currently capture. For instance, we had an intermediate **Difficult/Hard** code that captured general descriptions of challenge. This code's description relied narrowly on adjective use, e.g. "that class was impossible" or "that class was tough." We then imagined a student describing challenge in a more analogical way to something known to be intrinsically difficult, e.g. "that class was like climbing Mount Everest." This caused us to add clarifying language to this code to allow for the use of analogies like this.

As a final step to clarifying the language of our codes, we asked a new colleague (Nair) who had not seen our codes to become familiar with them and try to apply them to some student data. His use of our codes in practice brought up additional areas where the language of our codes was not entirely clear. We modified the language of our codes to address places of confusion. This process led to the finalized evidence and response codes presented in this paper.

C. Evidence for Challenge Codes Overview
In this section we present our evidence for challenge code names along with a short summary of the code, an idealized example, and a real student example from a selected set of physics course interviews described above. These are not the full codebook descriptions. The full codebook descriptions are included in a supplemental appendix. We break the codes into two main areas: activity difficulties and interactional difficulties. Note that these codes accept any magnitude of description in student dialogue, for example, "a little bit hard" and "very hard" are all accepted.

| Code Name | Code Description – Short Summary | Idealized Example Quote | Quote from Interview Set |
|---|---|---|---|
| *Activity Difficulties* | | | |
| Difficult/Hard | This code is the most general code, capturing any language that suggests difficulty. | That was a hard exam. | *The process to get that grade was very difficult* (Lian) |
| Task Failure | This code captures failure with respect to internal or external evaluation. This code allows for failure to occur on the way to success. | I wanted an A on that exam, but I got a C. | *We could not make that equation work out right* (Benjamin) |

| Complex/Confusing/Dense | This code captures conceptual difficulty. | Electricity and Magnetism just doesn't make sense to me | *I honestly thought all of physics was impossible to understand (Leyla)* |
|---|---|---|---|
| New/Low Background | This code captures activities, content areas, or learning strategies that are somewhat or entirely unfamiliar. | I've never done coding before. | *I didn't know how to use any of those tools (Leyla)* |
| Tedious | The code captures elements of an activity being repetitive or monotonous. | We had to solder all those little pieces to the breadboard and it took forever. | *And then, even the coding. Even though it's really tedious… (Leyla)* |
| Surprise at Success | This code captures an inferential measure of difficulty: an activity was difficult because a student was unsure about their ability to complete it successfully. | I never thought we'd be able to figure out that problem, but we kept at it and got it eventually. | *I was like, I can't believe I actually did this well on one physics exam. (Leyla)* |
| Failure/Struggle in Past Linked to Present | This code captures moments where students link past failure or struggle before the course to current descriptions of the course or an activity within it. | In high school I failed some exams, so I was scared about exams in this class. | *I always thought… I'm not really good at math. I've never been really good at math. So I didn't think I'd really enjoy physics very much (Maya)* |
| *Interactional Difficulties* | | | |
| Social Block/Difficulty | This code captures evidence that a student wants to reach out to another person for help or collaboration, but is blocked from doing so for any reason. It also captures any description | My lab group doesn't get along very well | *When I didn't do as well as I wanted on that [exam]…[the professor] was like, "Oh well, that's a good grade." And like, well, it's* |
| | that a social interaction is not ideal[2]. | | *not a good grade for me…So then I was kind of like… not disgruntled, but it didn't really make me want to do that much better. Like if you can't really help me or tell me what else I need to do, like I'm probably going to lose motivation in your class. (Anna)* |
| Peer Comparison - Worse | The code captures student descriptions of perceiving that one or many students are faster or better with respect to course activities, content, or overall grades. | I'm always the last one to hand in my exam. | *some of my peers…they'll see a problem, they just know how to do it, and I don't get know how they can do that (Leyla)* |

### D. Racism, Sexism, and Other Equity Issues in Student Accounts of Challenge

As we were developing our coding scheme, we recognized that some challenges were related to inequities connected to students' identities. At first, we developed a "marginalization" code to broadly capture student dialogue that dealt with descriptions of ableism, racism, sexism, and other -isms affecting students' interactions with their peers and instructors. We also noticed moments where an activity itself had problematic aspects to it that played a role in marginalization. For instance, one student talked about how physics problems sometimes involve stereotypically masculine activities such as throwing a football and how that advantages people who have performed those activities:

Anna*: It's hard sometimes, you'll be like, "Well, I've never thrown a football." Like I know it's supposed to go in that arc, but I've never thrown it in that arc, so it's like mine are very much not arc-y. So sometimes things are very hard to visualize in terms of, you know, if I throw it*

---

[2] Note that racism, sexism, ableism, xenophobia, etc., as well as implicit bias play powerful discriminatory roles in social interactions. We will address this partially in section VI.D, "Racism, Sexism, and Other Equity Issues in Student Accounts of Challenge." This code captures any social difficulty that may or may not have elements of marginalization.

*faster what is it going to do? I think that some of the guys who played football or who played football in the yard or whatever, they're like, "Oh, if I hurl it at somebody it's going to take a different path," or something like that.*

Upon reflecting on the idea of a marginalization code, however, it stuck out to us as ontologically different than other difficulty codes. Racism, as just one example, is a systemic problem, an "ingrained feature of our landscape" (Delgado & Stefancic, 2000). It is everywhere in U.S. culture and plays out in particular ways at primarily white universities (Harper et al., 2013; Harwood et al., 2015; Northwestern University, Black Student Experience Task Force, 2016) that comprise the context of our interviews. Therefore, we do not place student experiences of racism and other –isms alongside other difficulties. Nonetheless, issues of marginalization are important parts of STEM learning. In our limitations section, we return to the question of how to capture marginalization when studying mindset. For now, we note a theme that we did not explicitly create a difficulty code for, but that is critically important to understanding student experiences with physics course challenges.

E. Discussion

*1. Evidence for Challenge Codes in Comparison to the Mindset Literature*
We now return to our question: to what extent is evidence for challenge in college physics reflective of evidence for challenge found in the mindset literature?

Recall that our coding scheme relies on student descriptions of challenge rather than researcher-imposed challenges typical of the mindset clinical lab studies. This leads to some evidence for challenge that is distinct from the mindset literature. For instance, in physics class, what constitutes **Task Failure** is not always straightforward. A student working on a homework problem and seeing that their answer is wrong according to the back of the book is a straightforward example, however, students frequently set personal goals for themselves and can fail at those goals. For instance, a student might set a goal of achieving the grade of an A on an exam and instead receive a B. Thus, our **Task Failure** code is different from the mindset literature in that it can subjectively include students' own personal descriptions of failure. Similarly, the **Surprise at Success** code inherently brings personal history and expectations into one's experience in college physics. Surprise at success is frequently linked to a student's prior history with performing poorly in the past.

Interactional difficulty codes reflect a major difference between clinical lab studies with individuals and physics courses where many of the activities are collaborative in nature. Difficulties associated with peer and instructor interactions are lacking in the mindset literature. Mindset studies create a kind of artificial social block where research participants perform tasks alone. Students in transformed physics classes are rarely so isolated.

The **Tedious** code also reflects a major difference between challenges in clinical lab studies and in physics courses. When physics courses emphasize disciplinary skills such as coding and give students project-based experiences, elements of such work can certainly feel tedious. This is different from the constraints placed on clinical lab studies where it is difficult to ask participants, particularly children, to work on problems for more than an hour or two at a time.

Our finalized codes suggest the following result: college physics course challenges are sometimes significantly different from challenges previously described in the mindset literature, particularly when those challenges are interactional, tedious, and/or related to students' personal goals and histories. In addition, our codes capture a wider range of magnitude of challenge.

*2. Evidence for Challenge Codes: Mindset Threat Conditions*
Mindset researchers argue that one only sees bifurcation into fixed and growth behavior in the case that a challenge is threatening enough (Dweck, 2000). Kids with fixed and growth mindset beliefs (as measured by belief surveys) performed similarly on problems that were at their grade level. It was only when faced with problems above their grade level that researchers saw the split into two distinct kinds of response to challenge.

The "mindset threat" condition has been helpful to the clinical studies of mindset as it allows one to treat all challenges that are threatening enough in the same way theoretically. However, consider an attempt at characterizing a student's description of challenge as implying that they are worried enough about failure that it meets the mindset threat condition. First, this is methodologically difficult. There are cultural dynamics at play in how willing students might be to admit that they worried they might fail. Still, close attention to culture might allow researchers to argue that students are indeed worried or not worried that they might fail.

Second, a question arises on whether we should treat challenges in a binary way as either meeting a mindset threat condition or not meeting such a condition. There may be challenges where students are afraid of failing because they perceive their peers and instructors to be racist and this impedes their ability to connect to a classroom support network. The underlying theory for why a student might respond to a challenge negatively in such a case should be more related to racism than beliefs (see additional arguments made by McGee and Bentley, 2017). Psychological studies have sometimes treated growth mindset beliefs as buffering students from the negative impacts of gendered stereotypes in college math classes, for instance (Good et al., 2012). However, researchers have not returned to the definition of what constitutes a challenge to consider ways that students might view physics activities as more or less challenging in relation to interactional forms of marginalization.

Our description of our coding scheme development raises questions about a key aspect of the mindset literature: considering challenge as "threatening enough" or "not threatening enough." We argue that a finer grainsize description of challenge is often warranted to properly interpret students' responses to challenge. Our definition of challenge will necessarily impact how we interpret a student's response.

**VII: Examining Response to Challenge**

A. *A Priori* Codes from the Literature: Response to Challenge
A large area of the mindset literature is focused on characterizing research participants' responses to challenge as aligned with either a fixed or a growth mindset. These characterizations were built mainly in the context of examining participant behavior in-the-moment when experiencing challenge. Our goal was to characterize students' response to less contrived or more real-world challenges in the context of college physics. Therefore, we adapted these characterizations to end-of-semester interview dialogue.

In our methodology, we began by adapting the key elements of response to challenge from the mindset literature: *effort, strategy, emotion*, and *self-capability statements* (Dweck, 2000).

*Effort*
The most straightforward way that the mindset literature tracked response to challenge has to do with giving up versus putting in effort. In developing our coding scheme, we began with a priori response-to-challenge codes that involved a range of effort: **no effort, some effort, to high effort.**

*Strategy-Use*
Another way that the mindset literature tracks response to challenge has to do with employing or changing strategies. For instance, a standard study in the mindset literature involves giving children a few easy problems followed by problems that are much more challenging. In these studies, some children stopped employing strategies that they had previously used on easier problems (Dweck & Leggett, 1988). Dweck notes that some kids even "taught themselves new and more sophisticated strategies for addressing the new and more difficult problems" (Dweck, 2000). In our starting codes, we looked for problem-solving or metacognitive strategies. We included codes that tracked when students used a particular problem-solving strategy or reflected on their current strategies. We also included a code to tag places where strategies were clearly new or a change from prior work. Our starting codes were **problem-solving/metacognition** and **new/change problem-solving/metacognition**. Starting out, we thought it might be possible to break these codes into more specific ones.

*Emotion*
In Dweck's (2000) book, she describes "notable negative affect" associated with a helpless response to challenge. She notes that kids "began to express a variety of negative emotions" when they were met with difficulty, including boredom and despair. In the mastery-oriented response to challenge, kids "maintained a positive mood." Dweck notes cheerfulness and statements of excitement such as, "I love a challenge" (pg. 10). Following this literature, we developed a set of a priori emotion codes. These involved **positive emotions** and **negative emotions** generally as well as **liking/loving** or **hating/disliking** an activity.

*Self-Capability*
In the mindset literature, researchers describe a helpless or mastery-oriented response to challenge as involving optimistic or pessimistic self-capability statements (Dweck, 2000). Pessimistic statements would involve doubting one's intelligence, "denigrating" one's abilities, and blaming one's intelligence for failures. Kids would say things like, "I guess I'm not very smart," or, "I'm no good at things like this" (Dweck, 2000). Dweck notes that kids would become pessimistic about future success as well. Using these aspects of the literature we developed a set of a priori codes around **positive self-capability** and **negative self-capability** in both the present as well as looking toward the future.

Dweck (2000) also notes that kids sometimes made self-motivating or self-monitoring instructions like, "The harder it gets, the harder I need to try," (pg. 9). To account for this type of reflective statement, we added one additional a priori code to look for **meta-statements linking hard work or strategy-use to success**. Likewise, Dweck notes that kids sometimes made "deflection" statements where they discussed their success in other realms while performing poorly on logic problems (Dweck, 2000). This caused us to create a **deflection** category.

Note that, in the mindset literature's studies, kids sometimes make statements that sound a lot like beliefs. "I guess I'm not very smart" sounds very nearby some of the mindset belief survey items such as, "Your intelligence is something about you that you can't change very much" (Dweck, 2000). However, the mindset literature only measures beliefs through the use of Likert scale survey items. Self-capability statements made by research participants in response to challenge are not categorized as beliefs in the current psychological research on mindset. For the purposes of our coding scheme that helps us to understand the applicability of the mindset literature's approach in college physics, we follow the existing literature in paying attention to self-capability statements as a kind of response to challenge. The theoretical framework that underlies our approach (Elby & Hammer, 2010) suggests that these kinds of statements can be seen as in alignment with a more context-dependent approach to beliefs.

B. Modifying and Adapting Response to Challenge Codes: Selected Moments
The process of modifying and adapting our Response to Challenge codes followed a similar process to what is described in detail in our Evidence for Challenge codes methodology discussion. One major difference between the two coding development processes was that the mindset literature has more explicitly described how certain kinds of response to challenge are operationalized. This meant that we started with more specific a priori codes. Then, we followed a similar process to modification that was followed for our evidence for challenge codes. To reiterate: First, we examined student dialogue and this led us to modify details of our a priori codes and arrive at a set of intermediate codes. We then considered hypothetical examples to continue to refine our intermediate codes. Lastly, author (Nair) became familiar with the codes and tried to apply them to some student data. We modified the language of our codes to address places of confusion. This process led to the finalized evidence and response codes presented in this paper.

During the code modification process, some of our codes became more general in nature than the mindset literature's versions of coding. This was due to the complexity and idiosyncrasy of student dialogue about transformed college physics courses with project-based elements. For instance, in the mindset literature, the kinds of tasks given to participants made it more possible to track strategy-use more specifically. In one study by Deiner & Dweck (1980), researchers gave children logic problems. They classified children's strategies as "useful" or "ineffective." Furthermore, within "useful" strategies, researchers had multiple specific strategies such as "hypothesis checking" that were coded and ranked in order of their efficiency for solving problems. This allowed researchers to make statements such as "more than two thirds of [children] showed a clear decline in the level of their problem-solving strategy under failure and over 60% lapsed into ineffective strategies…that would never yield a solution" (Dweck & Leggett, 1988). In end-of-semester interview data, it is difficult to assign gradation to strategy-use. We thus took an approach of classifying strategies more generally as either related to a course activity or related to reaching out to other people.

Reaching out to other people involved a new set of codes that involved interactions with others. We noticed student descriptions of working with other people or attending office hours. When we first noticed examples along these lines, we had a broad category that tagged any moment of interaction. However, we decided to break apart general descriptions of working with peers in a study group appeared different than **Seeking Out Help** (not peers) and **Working with Peers**. We made this distinction because explicitly seeking out help outside of one's study group appeared to demonstrate an extra level of being willing to vulnerably tell other people about one's lack of understanding. For instance, Dweck (2010) notes that a fixed mindset often results in students working to "hide their deficiencies." We also recognized that some response to challenge codes appeared to be able to do dual work for us in that they could be considered evidence for challenge. By splitting apart study-group work from explicitly seeking out help, the **Seeking Out Help** code became a code that could be considered evidence for challenge. Mention of working in a study group could not be as easily considered direct evidence of an activity being challenging since some students work in study groups whether or not they find the homework challenging.

Drawing on the mindset literature, we originally began with three effort categories: none, some, and high. We quickly realized that almost every activity that students described in their interviews involved at least some effort. This may particularly be the case in our context: a transformed college physics class with predominately pre-med students. We found ourselves applying the "some effort" code so many times (often every line of transcript had some evidence of some effort) that it became a meaningless code. For instance, "I worked on the homework" would be evidence of some effort. This caused us to more clearly articulate what was meant by high effort and we arrived at the **Hard Work/Significant Time** code. Our interviews did not have instances of lack of effort. However, one student mentioned that he would have given up early before fully completing his work under certain circumstances. This caused us to create a more meaningful "some effort" code called **Giving Up/Stopping**. We could imagine giving up on work when feeling a low sense of self-capability and so this code seemed important to consider. Of course, explanations beyond self-capability, such as having other courses to balance, are possible, and will be explored in the discussion of these codes.

C. Some Challenge Codes Dually Act as Evidence of Challenge
In the process of finalizing our response to challenge codes, we recognized that some codes could be dually considered evidence for challenge. In fact, some of our finalization process involved clarifying codes so that we could use some of the codes as dual-evidence. Student descriptions fitting into the **Hard Work/Significant Time** code, for instance, seemed at face value to convey at least some level of challenge. We looked at each response code and asked, "could every possible example that this code covers also be considered evidence for challenge?" Some codes were more mixed. For instance, the **Negative Future** code describes "any range of dialogue that suggests they do not want to engage in more physics courses or disciplinary practices in the future." There might be some versions of dialogue picked up by this code that suggest challenge. However, we could also consider many reasons why a student might mention not wanting to take physics again. For instance, they might have too full of a course schedule or not feel that physics is related to their major. Members of our research team (Little, Humphrey, and Green) examined each response code and considered whether all examples covered by the code could be considered to count as evidence for challenge. Five of our codes passed this test and we suggest that these codes can be dually considered to be both evidence for challenge and response to challenge: **Hard Work/Significant Time**, **New/Change – Strategies General**, **Seeking Out Help (not peers)**, **Dislike/Hate**, and **Discomfort**. These codes can provide additional evidence for challenge. Student dialogue is not always explicitly descriptive in a way that fits our evidence for challenge codes; thus, the response codes that can be dually used as evidence for challenge do important work in characterizing student dialogue.

D. Response to Challenge Codes Overview

In this section, we present our response to challenge code names along with a short summary of the code, an idealized example, and a real student example, if it exists, from a selected set of physics course interviews described in the data collection section. These are not the full codebook descriptions; as they are considerably longer, the full codebook descriptions are included in an appendix. We break the codes into four main areas: effort, strategy use, emotion, and self-capability. In the final column, we note whether the code aligns with the mindset literature's description of a growth or fixed mindset response to challenge. We will explore alternative explanations to fixed or growth mindset interpretations in our discussion.

| Code Name | Code Description | Idealized Example Quote | Quote from Interview Set | GM or FM? |
|---|---|---|---|---|
| *Effort* | | | | |
| Hard Work/ Significant Time | The student describes the work that they put into an activity as significant in some way | That homework set took us forever to finish | *I really tried to study hard on that exam* (Lian) | GM |
| Giving Up/Stopping | The student describes stopping before an activity is complete for any reason | I didn't have enough time finish the homework | Does not exist in our data set | FM |
| No Time/Avoidance | The student describes putting no time at all into an activity for any reason | I couldn't make myself sit down to study for that exam | *I just decided not to [do it]…because…it's a lot to learn and I don't have the time* (Jackson) | FM |
| *Strategy Use* | | | | |

| Strategies-General | This code accepts all kinds of strategies from problem-solving to metacognition except for the strategy of reaching out to talk to other people. Strategies can be described in detail or only vaguely but still count. | I always draw a Free Body Diagram first when I'm solving mechanics problems | *I was using my lab report to study for this last exam* (Leyla) | GM |
|---|---|---|---|---|
| New/Change – Strategies General | This is the same code as described above, but there must be evidence that the strategy use is either completely new or an intentional change. | I was struggling on the exams, but a friend taught me a different way to study that really helped | *The one thing that I thought wasn't important to study was the most important thing to study. So now that I switched my order, it's a lot better.* (Julie) | GM |
| Seeking Out Help (not peers) | In this code students describe getting help by talking to any person except for a physics class peer. | I went to my faculty member's office hours | *I went to help rooms [where tutors are available]* (Lian) | GM |
| Working with Peers | This code captures any mention of working with a physics class peer for any amount of time, e.g. a study group. However, if an activity (e.g. a lab) has been required to | My friend and I always meet up to do the homework together | *[My friend and I] study for all of our exams together and stuff and we just argue all the time…* (Kasee) | -- |
| | be performed in a group by an instructor, it does not count. | | | |
| Ineffectual Strategies | This code captures students describing the use of strategies that will clearly not move them forward toward completing a task effectively. | I just started writing down random physics equations and plugging in numbers so at least I had something written down. | Does not exist in our data set | FM |
| *Emotion* | | | | |
| Dislike/Hate | This code focuses on student dialogue where they use words and phrases such as dislike, hate, don't like, or don't love. | I didn't really like thermodynamics. | *I think I like more, the lecture, and then practice problems, rather than just getting thrown into labs…I liked some of it, and then <u>some of it, I didn't like</u>* (Leyla) | FM |
| Like/Love | This code focuses on student dialogue where they use like or love explicitly. | I love this class. | *I actually like this style of learning a lot better* (Benjamin) | GM |
| Positive Emotion | This code captures student dialogue about positive emotions, either commonly accepted U.S. emotions and/or a statement | I felt so excited about how good our project was. | *I was <u>ecstatic.</u> I was like, "I can't believe I actually did this well on one physics exam."* (Leyla) | GM |
| | about feeling or a state of feeling, e.g. I'm happy or I felt happy. | | | |
| Emotional Discomfort | This code captures emotions similarly to the above category, but focuses on discomfort which students do not explicitly link to any negative repercussions. | I was scared we wouldn't finish the homework | *It's always really stressful going into [exams]* (Benjamin) | FM |
| Lessening of Emotional Discomfort | This code captures student descriptions of the lessening of emotional discomfort | I'm less scared of physics than when I started | *It just feels a lot better* (Benjamin) | -- |
| Positive Future | This code captures any range of student dialogue that suggests they want to engage in more physics courses or disciplinary practices in the future. | I'm going to try to take an elective physics course next year. | *Even though [coding is] really tedious, and it's sometimes hard…I wish I could somehow incorporate that into my job as a surgeon, somehow.* (Leyla) | GM |
| Negative Future | This code captures any range of dialogue that suggests they do not want to engage in more physics courses or disciplinary practices in the future. | I never want to see another physics book ever again | *I used to want to do astrophysics, and then I did that [summer camp] and I was like, "Nope, this is not for me."* (Kasee) | FM |
| *Self-Capability* | | | | |

| Positive Smart Label | This code captures student dialogue where they give any kind of smart label to themselves as a person rather than their actions. | I've always been naturally good at physics. | *Yeah, I'd say I'm a physics person [after taking this physics class].* (Kasee) | -- |
|---|---|---|---|---|
| Negative Smart Label | This code captures student dialogue where they give any kind of not-smart label to themselves as a person rather than to their actions. | I'm terrible at this class. | *That's why I think I'm bad at physics* (Leyla) | FM |
| Unable To | This code captures a student describing their lack of ability to complete a goal or their lack of ability with physics disciplinary skills. | I can never figure out what's wrong with our code | *we could not make that equation work out right* (Benjamin) | FM |
| Able to | This code captures student success in achieving a goal or a sense that they now have a capability that they did not previously have. | My goal was to get a B on that exam and I did it. | *I can look at a light, and I understand what's happening* (Leyla) | GM |
| Better/Improvement | This code captures improvement where students do not mention | I've gotten faster at taking exams. | *It was the highest [grade] I've ever…gotten, I think* (Leyla) | GM |

| | | | success or being able. | |
|---|---|---|---|---|
| Deflection | This code captures any mention of skills or strengths outside of the domain that one is telling a story of struggle within. | I'm not good at this class, but I just won an award. | *So it was kind of fun to have [an English class]…but…I didn't think the way a lot of the English students did, and that was kind of difficult. Yeah, I thought a lot more scientifically and a lot more on the surface… like their analysis would just be so much deeper, and just kind of like the way I would think… was just a lot more scientific and straightforward[3].* (Maya) | FM |
| Meta Statement – Importance of Effort | This code captures any statement where students link effort or strategy-use to success or lack thereof. | I didn't do well on that exam, but that's because I wasn't using the right strategies. | *But I believe if I do put more time into it I will understand* (Lian) | GM |

## E. Discussion

<u>*1. Response to Challenge Codes in Comparison to the Mindset Literature*</u>
We began with the question, "can we see responses to challenge in college physics that resemble that of the mindset literature?" Our coding scheme does allow us to find responses to challenge in college physics that broadly fit into the mindset literature's four areas of responses: effort, strategy-use, emotion, and self-capability statements.

Our response to challenge codes were developed in the context of transformed college physics courses and therefore capture a somewhat different range of responses than are found in clinical lab studies. We highlight some of these differences and the difficulties they raise for a mindset interpretation of students' responses to challenge, i.e. that beliefs about intelligence influence students' responses. We take these codes at face value and ask: are there some obvious factors that might supersede intelligence beliefs as explaining why a student is responding in that way? In the following section, we present some of our largest concerns for an intelligence-beliefs-related interpretation.

---
[3] Note that this is a deflective statement in the context of Maya's English class rather than in the context of a physics class. We kept the examples in the context of physics class wherever possible.

Effort
**Giving Up/Stopping** in particular raises concerns about a mindset interpretation of this response. For instance, in one of our interviews, a student mentioned strategically putting time into courses. They noted that if their two physics midterms scores made it of low likelihood that they could get an A in their physics class, "then I wouldn't really try as hard as right now for the past two exams… Then I'm going to put more of my time into other class that I could have a potential to get an A." Clinical lab challenges do not have these more difficult decisions embedded. Dweck and colleagues (1988) do not ignore that people must make hard choices sometimes: "of course, individuals need to be able to gauge when tasks should be avoided or abandoned," however, most of the studies of challenge in the mindset literature minimize the complexity of this choice. This cannot be ignored in the context of college physics.

Strategy-Use
In terms of a mindset interpretation for our codes, there are parallels to make with the Effort codes. Much like the **Giving Up/Stopping** effort code, the reasons behind the **Ineffective Strategies** code may be similarly varied. Students may use ineffective strategies if they run out of time on an activity, not necessarily because they do not believe it is possible to complete the activity or that it is a "threat to their self-esteem" (Dweck & Leggett, 1988).

Emotion
With respect to a mindset interpretation of these emotion-related responses to challenge, the emotional landscape becomes more complicated in college physics than in traditional mindset studies. Physics can be frustrating at times and we do not expect students to necessarily maintain a "positive mood" (Dweck, 2000) at all times during challenges. The challenges we examine are frequently much longer in timescale than a psychology lab logic problem set activity. A weeks-long project can have moments where students' emotions will reasonably vary. As Jaber & Hammer's (2016) work on epistemic affect suggests, even children may reasonably experience a wide variety of emotions when engaged in doing science such as, "fascination, curiosity, frustration, boredom, surprise, and so on." Emotions that might, on their face, seem negative, can ultimately lead to positive outcomes. For instance, "student frustration and struggle" are "necessary features of learning environments that promote ownership" (Dounas-Frazer & Lewandowski, 2017). Therefore, our a priori codes from the mindset literature involving negative emotions evolved to a less value-laden final code of **Emotional Discomfort**. Is a feeling of boredom arising because of one's beliefs about intelligence or because a science activity is becoming tedious? Both are plausible interpretations of a student expressing boredom. Likewise, is a feeling of fear expressed by a student about their exam arising because of beliefs about intelligence or because of test-taking anxiety? It is unclear what the relationship might be between student beliefs and **Emotional Discomfort**, which could be caused by a variety of factors.

Self-Capability

The **Positive Smart Label** arose from our data and is outside of how the mindset literature categorizes self-capability in response to challenge. As an example from one of our data sets, one student found statistics challenging but noted that it "*definitely feels like it's something I'm good at, and it feels pretty intuitive. I joke with people in the class that it's genetic...*" Researchers in the mindset literature might suggest that the magnitude of challenge is not high enough in moments where we see positive smart labels. However, with respect to the range of challenges in college physics, positive smart labels are possible.

Our **Unable To** code raises concerns about a mindset interpretation of this response to challenge. Recall that in the mindset literature, fixed mindset self-capability statements had to do with questioning or condemning one's intelligence. The idea was that when a task became so challenging such that failure was likely, growth mindset students would cheer themselves on while fixed mindset students would give up and question their intelligence. In student interview data, we pulled apart mention of ability from categorical personal labels. We noticed that students sometimes spoke to their ability or inability to succeed or enact a skill, e.g. "*we could not make that equation work out right*." Mention of inability was frequently transient, followed by accounts of ultimately succeeding. Therefore, mention of inability cannot necessarily be interpreted as condemning one's intelligence.

Lastly, our **Unable To/Able To** as codes can be seen as places where mindset has connections with the construct of self-efficacy (Bandura, 1977). As Sawtelle et al. (2012) point out, "Experiences with successful completion of a task should have a strong positive influence on an individual's confidence in their ability to complete a similar task." Thus, when students face a challenge about understanding circuits and lightbulbs and can ultimately say, "I can look at a light, and I understand what's happening," this could be considered a self-efficacy building moment. Our team is currently exploring connections between self-efficacy and mindset. These connections are backgrounded in this paper but will be explored in future work.

*2. In-the-Moment Challenge vs. End-of-Semester Interview Dialogue*
Much of the mindset literature's characterization of response to challenge occurs in-the-moment in clinical lab studies. Our study involves something quite different: end-of-semester descriptions of challenge. End-of-semester interviews mean that students can talk about a range of content-areas and activities. This does introduce more complexity into what might be informing a student's response to challenge. We see some of that complexity in our response to challenge codes and our discussion of the interpretation of these codes.

Is it possible to minimize this complexity? There is one activity in college physics classrooms that appears to be the closest approximation to mindset-type challenges: a student taking a physics exam. If an exam is taken individually and if the exam is viewed by students to be confusing and induces some fear of failure, this begins to sound like our a priori mindset challenge codes: isolated, complex/confusing, and some likelihood of task failure. However, even here, in what might be considered the least complex of college physics challenges, complexities arise. What if a student experiences test-taking anxiety (Perry, 2004) or stereotype threat (Steele & Aronson, 1995)? What if a student has a negative relationship with the faculty member proctoring the exam where that faculty member has expressed doubts about their abilities? This thought experiment suggests that we are "stuck" with complexity if we truly want to understand mechanisms behind the ways that students respond to challenge in college physics. Therefore, we must embrace methods and frameworks that give us footholds to understand the richness of how students experience and respond to challenges in transformed and project-based college physics classrooms.

**VIII. Limitations**
Our coding scheme development suggests that individual belief-focused theoretical frameworks for mindset research, even ones that allow for more context-dependent variation, have their limitations. We found that not all challenges are alike and attending to the variety of factors at play in what makes an activity challenging is necessary for making interpretations of student responses to challenge. We posit that existing research on science identity may provide useful theoretical frameworks and methodologies for mindset researchers. For instance, in parallel to studies on identity, one could take a lens of examining classroom activity structures that may make some responses to challenge more or less available to students in certain moments (Shah, 2013; Carlone et al., 2014; Hand & Gresalfi, 2015). Given physics education research demonstrating that many physics faculty express ideas around physics ability as innate (Zwickl et al., 2014; Dounas-Frazer & Lewandowski, 2017; Scherr et al., 2017), examining the classroom and departmental environments that faculty set up should be a focus of study. For instance, Johnson et al. (2017) have suggested that faculty messaging that aligns with growth mindset is one aspect of a departmental culture that is supportive of women of color.

Although we noted racism and other forms of marginalization as playing a role in physics course challenges, we have not presented a methodology for taking this into account. In addition, our interviews were primarily with white students such that experiences with racism were not a central feature of our data. Again, identity research can guide us; Hyater-Adams et al. (accepted for publication, 2018) have built methodological tools to look at "identity at an internal, interpersonal, and institutional level, with special attention to the intersections of racial and physics identities." Research explicitly examining these intersections with respect to mindset is needed.

Lastly, our coding scheme was developed with students who skewed toward more positive and successful experiences with college physics. A focus on students who have negative and less successful experiences may suggest additional categories of importance in understanding students' responses to challenge.

**XI. Conclusion**
Very little is actually known about the processes that impact students' responses to challenge in the context of transformed college physics courses. The kinds of challenges found in college physics, particularly those that are interactional, tedious, and related to students' personal goals and histories, have not been the focus of study in the mindset literature. What does our coding scheme development and data analysis suggest about ways of moving forward as a field?

We have responded to Schwartz et al.'s (2016) call for better understanding of, "causal forces that could never appear in a laboratory." For instance, in talking with college physics students who sought to go onto medical school, we saw a complication to mindset lab studies. Students hoping to achieve high grades may not easily give up, even if they ascribe to fixed-mindset aligned intelligence beliefs. As we saw with Leyla at the beginning of this paper, students can describe seemingly-conflicting ideas about working at and becoming capable in physics, while still identifying with "bad at physics" labels. Therefore, we expect college physics courses with predominately life sciences student populations to be particularly complex and insightful terrain with respect to mindset.

We have provided the first coding scheme that allows researchers to look more closely at student dialogue toward understanding challenges students might face in introductory physics classes. Our coding scheme has broad utility: for those seeking to understand mindset's applicability to college physics and for those attempting to characterize student experiences of challenge more generally. For those seeking to apply a more context-dependent belief framework in interpretation of student dialogue and action, our mindset response codes can aid in such work. This helps the field to move past broad Likert-scale survey measures of beliefs.

**Acknowledgements**
We acknowledge Lyman Briggs College, the Science and Society @ State (S3) Collaborative Grant, and the STEM Gateway Fellowship through the Create4STEM Center at Michigan State University for supporting this work. We thank Ben Geller, Gina Quan, and Dimitri Dounas-Frazer for their thoughtful feedback on this paper. We thank Catherine Crouch, Ben Geller, and Chandra Turpen for sharing interview data that informed our coding scheme development.